# Frequency-domain probe beam deflection method for measurement of thermal conductivity of materials on micron length scale

Jinchi Sun,[1] Guangxin Lyu,[1] and David G. Cahill[1, a)]

[1]Department of Materials Science and Engineering and Frederick Seitz Materials Research Laboratory, University of Illinois, Urbana, Illinois 61801, USA

a)Author to whom correspondence should be addressed: d-cahill@illinois.edu.

Time-domain thermoreflectance (TDTR) and frequency-domain thermoreflectance (FDTR) have been widely used for non-contact measurement of anisotropic thermal conductivity of materials with high spatial resolution. However, the requirement of high thermoreflectance coefficient restricts the choice of metal coating and laser wavelength. The accuracy of the measurement is often limited by the high sensitivity to the radii of the laser beams. We describe an alternative frequency-domain pump-probe technique based on probe beam deflection. The beam deflection is primarily caused by thermoelastic deformation of the sample surface with a magnitude determined by the thermal expansion coefficient of the bulk material to measure. We derive an analytical solution to the coupled elasticity and heat diffusion equations for periodic heating of a multilayer sample with anisotropic elastic constants, thermal conductivity, and thermal expansion coefficients. In most cases, a simplified model can reliably describe the frequency dependence of the beam deflection signal without knowledge of the elastic constants and thermal expansion coefficients of the material. The magnitude of the probe beam deflection signal is larger than the maximum magnitude achievable by thermoreflectance detection of surface temperatures if the thermal expansion coefficient is greater than $5\times10^{-6}$ $K^{-1}$. The sensitivity to laser beam radii is suppressed when a larger beam offset is used. We find nearly perfect matching of the measured signal and model prediction, and measure thermal conductivities within 6% of accepted values for materials spanning the range of polymers to gold, 0.1 - 300 W/(m K).

## I. INTRODUCTION

Pump-probe optical techniques based on either time-domain thermoreflectance (TDTR) [1–4] or frequency-domain thermoreflectance (FDTR) [5] have been developed for non-contact measurement of thermal conductivity of thin film and bulk materials, and are widely used for measurement of anisotropic thermal conductivity [6–12] and thermal conductivity mapping with micron-scale spatial resolution [13–19]. In these techniques, a metal coating serves as the thermometer by generating a thermoreflectance signal; i.e., the signal derives from the change of reflectivity $R$ with temperature. Combinations of the metal and laser wavelength that give large thermoreflectance coefficient $\frac{1}{R}\frac{dR}{dT}$ and thus large magnitude of the signal are typically used to maximize the signal-to-noise ratio, for example, Au at a laser wavelength of 532 nm and Al at a laser wavelength of 785 nm [20]. However, for accurate measurements of in-plane thermal conductivity or high spatial resolution measurements of materials with low thermal conductivity, a metal with low thermal conductivity is needed to suppress heat spreading in the metal coating [21,22].

The accuracy of TDTR and FDTR is often limited by the systematic error propagated from the uncertainties in the pump and probe beam radii [23]. This problem is compounded when working at high spatial resolution because of the small depth of focus and the difficulty in accurately determining the intensity profile on small length scales. In a thermal conductivity mapping experiment, a drift in the position of the sample surface relative to the focal point of the laser beams propagates into systematic errors in thermal conductivity as a function of position.

In this paper, we describe a frequency-domain probe beam deflection (FD-PBD) method for measurement of thermal conductivity. This approach features larger magnitude of signal, flexible choices of the metal coating and laser wavelength, and smaller errors due to uncertainties in the beam radii. We have previously used a FD-PBD method in a back-side detection geometry to measure the thermal conductivity of transparent polymers attached to an Al-coated silica substrate [24,25]. In this previous version of FD-PBD, the polymer sample itself serves as the thermometer in the measurement by generating the probe beam deflection signal via the temperature dependence of the index of refraction and this technique is therefore limited to transparent and translucent materials. This approach is also limited to materials with small elastic constants so that thermal stresses in the material being studied do not significantly deform the silica substrate.

The FD-PBD approach we describe here removes these limitations. Basically, the bulk material to be measured is first coated by a metal film. During the measurement, the metal film is heated by the modulated pump beam. The thermal expansion of the material drives thermoelastic deformation of the sample surface and consequently deflection, i.e., a small change in direction, of the reflected probe beam. The probe beam deflection signal is measured as a function of the modulation frequency of the pump beam and the data are fitted to a model to extract the thermal conductivity of the material. The rest of the paper is structured as follows. In section II, the model needed is derived in detail. In section III, the experimental measurement of probe beam deflection is explained. In section IV, analysis of magnitude of the signal, discussion of simplification of the model, sensitivity and uncertainty analysis, and fitting results of the experimental data are presented. The model shows nearly perfect matching with the measured FD-PBD signal. The measured thermal conductivities agree well within 6% with accepted values in the range of 0.1 – 300 W/(m K).

## II. MODEL OF PROBE BEAM DEFLECTION

### A. Isotropic free thermal expansion model

This simplified model neglects all mechanisms of probe beam deflection except the deformation of the surface of the bulk material due to free thermal expansion, i.e. without mechanical constraint by the metal coating. The material is assumed to be a linear elastic semi-infinite solid with isotropic elastic and thermal expansion properties. The heating is due to absorption of a gaussian pump beam by the metal coating. The solution of probe beam deflection for this situation has been reported in our previous work on probe beam deflection [26,27], which is based on a solution of surface displacement derived with non-Fourier law of heat conduction equation that involves a relaxation time [28]. Here, we obtain the temperature field with Fourier law of heat conduction equation instead, while the derivation of probe beam deflection given the temperature field essentially follows previous work [26–28] and is elaborated here to fill in some intermediate steps in the derivation not included in the original paper [28].

Since the heating by pump beam has cylindrical symmetry, the temperature field $T$ and displacement field $\mathbf{u}$ also have cylindrical symmetry. The calculation can thus be greatly accelerated by selecting a cylindrical coordinate system and making use of Hankel transforms. As shown in Fig. 1a, we choose $r = 0$ at the center of the pump beam, $z = 0$ at the surface of the bulk material and $\hat{\mathbf{z}}$ pointing toward the interior of the sample. The displacement field $\mathbf{u} = u_r \hat{\mathbf{r}} + u_z \hat{\mathbf{z}}$ can be expressed by a dilatational potential $\varphi$ and a rotational potential $\psi$ with corresponding governing equations [28]:

$$u_r = \frac{\partial \varphi}{\partial r} + \frac{\partial^2 \psi}{\partial z \partial r}, \quad u_z = \frac{\partial \varphi}{\partial z} - \frac{1}{r}\frac{\partial}{\partial r}\left(r \frac{\partial \psi}{\partial r}\right), \quad (1)$$

$$\nabla^2 \varphi - \ddot{\varphi}/v_L^2 = \gamma T, \quad v_L^2 = (\lambda + 2\mu)/\rho, \quad (2a)$$

$$\nabla^2 \psi - \ddot{\psi}/v_T^2 = 0, \quad v_T^2 = \mu/\rho, \quad (2b)$$

where $\gamma = (3\lambda + 2\mu)/(\lambda + 2\mu)\, \alpha_T$, $\lambda$ and $\mu$ the Lamé parameters, and $\alpha_T$ the linear coefficient of thermal expansion. With the mechanical constraint by the metal coating neglected, Eq. (1) is subjected to the stress-free condition at the interface between metal coating and the bulk material,

$$\sigma_{zr}|_{z=0} = 0, \quad \sigma_{zz}|_{z=0} = 0. \quad (3)$$

Via Hooke's law, the strain-displacement relation [29] and Eq. (1), $\sigma_{zr}$ and $\sigma_{zz}$ in Eq. (3) can be expressed by $\varphi$ and $\psi$, giving

$$2\frac{\partial \varphi}{\partial z} + \frac{\partial^2 \psi}{\partial z^2} - \frac{1}{r}\frac{\partial}{\partial r}\left(r\frac{\partial \psi}{\partial r}\right) = 0, \quad (4a)$$

$$\frac{\partial^2 \varphi}{\partial z^2} - \frac{2v_T^2/v_L^2 - 1}{r}\frac{\partial}{\partial r}\left(r\frac{\partial \varphi}{\partial r}\right) \quad (4b)$$

$$- \frac{2v_T^2/v_L^2}{r}\frac{\partial}{\partial r}\left(r\frac{\partial^2 \psi}{\partial r \partial z}\right) - \gamma T = 0.$$

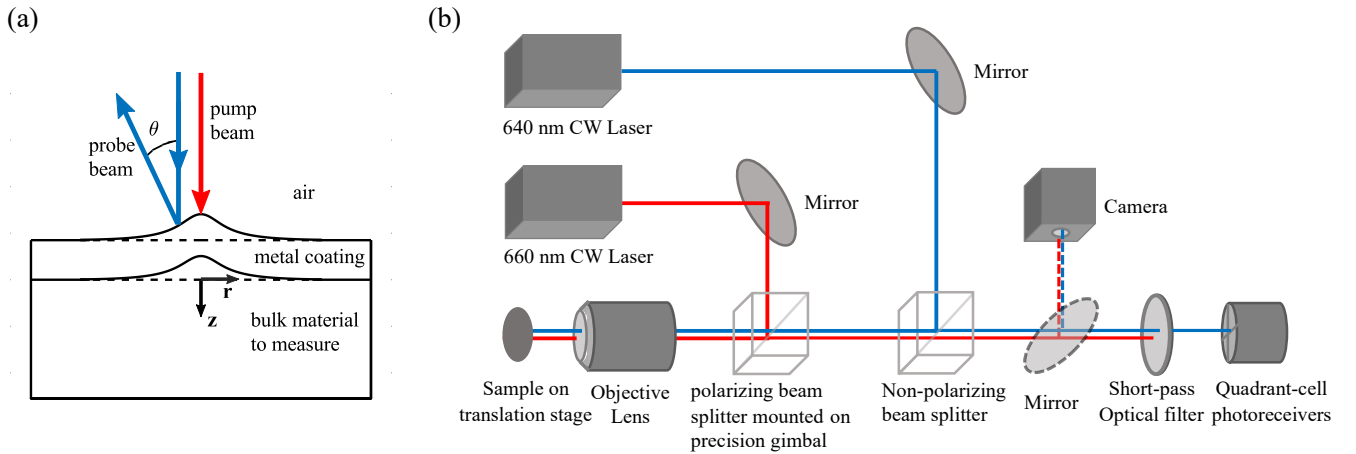

FIG. 1. Illustration of frequency-domain probe beam deflection (FD-PBD) system: (a) Sample geometry, (b) Optics. The pump beam (in red) and probe beam (in blue) are CW lasers at wavelength of 660 nm and 640 nm, respectively, with have an offset along vertical direction. The pump beam is modulated while the probe beam is unmodulated. A quadrant-cell photoreceiver is used with a lock-in amplifier to measure the probe beam deflection along vertical direction at modulation frequency of pump beam $f$. If $f$ is above the upper limit of measurement of the quadrant-cell photoreceiver, the probe beam is modulated at frequency $f + f_{dec}$ to measure the signal at the difference frequency $f_{dec}$.

After Hankel transform ($r$ to $k$) and Fourier transform ($t$ to $\omega$), the governing equations Eq. (2) become

$$\frac{\partial^2 \tilde{\tilde{\varphi}}}{\partial z^2} = \alpha^2 \tilde{\tilde{\varphi}} + \gamma \tilde{\tilde{T}}, \quad \alpha = (k^2 - \omega^2/v_L^2)^{1/2}, \quad (5a)$$

$$\frac{\partial^2 \tilde{\tilde{\psi}}}{\partial z^2} = \beta^2 \tilde{\tilde{\psi}}, \quad \beta = (k^2 - \omega^2/v_T^2)^{1/2}, \quad (5b)$$

while the boundary conditions Eq. (4) become

$$2\frac{\partial \tilde{\tilde{\varphi}}}{\partial z}\Big|_{z=0} + \frac{\partial^2 \tilde{\tilde{\psi}}}{\partial z^2}\Big|_{z=0} + k^2 \tilde{\tilde{\psi}}\Big|_{z=0} = 0, \quad (6a)$$

$$\frac{\partial^2 \tilde{\tilde{\varphi}}}{\partial z^2}\Big|_{z=0} + k^2(2v_T^2/v_L^2 - 1)\tilde{\tilde{\varphi}}\Big|_{z=0} \quad (6b)$$
$$+ 2k^2 v_T^2/v_L^2 \frac{\partial \tilde{\tilde{\psi}}}{\partial z}\Big|_{z=0} - \gamma \tilde{\tilde{T}}\Big|_{z=0} = 0.$$

The temperature field $\tilde{\tilde{T}}$ is governed by Fourier law of heat conduction after Hankel and Fourier transform [30]:

$$\frac{\partial^2 \tilde{\tilde{T}}}{\partial z^2} = \zeta^2 \tilde{\tilde{T}}, \quad \zeta = \left(\frac{\Lambda_r}{\Lambda_z} k^2 + \frac{i\omega C}{\Lambda_z}\right)^{1/2}, \quad (7)$$

where $\Lambda_z$ and $\Lambda_r$ are the cross-plane and in-plane thermal conductivity, respectively, and $C = \rho c_p$ the volumetric heat capacity, giving the temperature field in the bulk material

$$\tilde{\tilde{T}} = T_{bs} e^{-\zeta z}, z > 0, \quad (8)$$

where the surface temperature $T_{bs} \equiv \tilde{\tilde{T}}|_{z=0}$ is solved by transfer matrix approach for heat conduction in semi-infinite air/ metal coating/ semi-infinite material system as detailed in literature [2,30]. Basically, Eq. (7) applies to each layer given the thermal conductivity and volumetric heat capacity. Adiabatic boundary condition applies at $z \to \pm\infty$, while heat flux boundary condition applies at the air/metal interface with the heat flux equals

$$\tilde{\tilde{P}} = P_0 e^{-w_0^2 k^2/8}, \quad (9)$$

where $P_0$ is the amplitude of the fundamental harmonic of the absorbed power of modulated pump laser, $w_0$ the $1/e^2$ radii of pump beam.

With temperature field given by Eq. (8), the general solution of the governing equations Eq. (5) is:

$$\tilde{\tilde{\varphi}} = \gamma T_{bs}\left(a_\varphi e^{-\alpha z} + \frac{e^{-\zeta z}}{\zeta^2 - \alpha^2}\right), z > 0, \quad (10)$$

$$\tilde{\tilde{\psi}} = \gamma T_{bs} a_\psi e^{-\beta z}, z > 0, \quad (11)$$

where the constants $a_\varphi$ and $a_\psi$ is uniquely determined by the boundary conditions Eq. (6) to be

$$a_\varphi = -\frac{\zeta}{\alpha(\zeta^2 - \alpha^2)} + \frac{\beta^2 + k^2}{2\alpha} a_\psi,$$

$$a_\psi = \frac{2v_L^2\left(\alpha^2 + k^2(2v_T^2/v_L^2 - 1)\right)}{v_T^2(\zeta + \alpha)((\beta^2 + k^2)^2 - 4k^2\alpha\beta)}.$$

The surface displacement of this "isotropic free thermal expansion" model, $\tilde{\tilde{Z}}_{iso-free}(k,\omega) \equiv -\tilde{\tilde{u}}_z|_{z=0}$ (the minus sign makes displacement outwards positive) is given by

$$\tilde{\tilde{Z}}_{iso-free}(k,\omega) = \frac{\beta^2 - k^2}{2} \gamma T_{bs} a_\psi. \quad (12)$$

The "low frequency approximation" is applicable is most cases,

$$\omega^2/v_T^2 \ll k^2, \quad \omega^2/v_L^2 \ll k^2. \quad (13)$$

To see this, given a radii of pump beam $w_0 \sim 10$ μm, consider a relatively extreme case of frequency as high as $f = \omega/2\pi \sim 1$ MHz and speed of sound as low as $v_T, v_L \sim 1000$ m/s, then $\omega^2/v_T^2, \omega^2/v_L^2 \sim (2\pi/1000$ μm$)^2 \ll (2\pi/w_0)^2 \sim k^2$. Note that the frequency of interest is typically smaller than 1 MHz because the frequency of interest shifts to smaller values with decreasing thermal diffusivity (as shown later) and 1 MHz is for a high thermal diffusivity material such as gold.

Making use of Eq. (13), Eq. (12) reduces to

$$\tilde{\tilde{Z}}_{iso-free}(k,\omega) = \frac{(3\lambda + 2\mu)/(\lambda + 2\mu) \alpha_T}{1 - v_T^2/v_L^2} \frac{\tilde{\tilde{T}}_{bs}}{k + \zeta} \quad (14)$$
$$= 2(1+\nu)\alpha_T \frac{T_{bs}}{k + \zeta},$$

where the second equality is via the conversion formulae between elastic properties and $\nu$ is the Poisson's ratio. In this low frequency approximation, the frequency dependence of the surface deformation on elastic properties is eliminated, which is desirable for frequency-domain measurement of thermal conductivity, heat capacity, and coefficient of thermal expansion.

An inverse Hankel transform is then applied to obtain the surface displacement in real space,

$$\tilde{Z}_{iso-free}(r,\omega) = \quad (15)$$
$$\frac{1}{2\pi}\int_0^{+\infty} \tilde{\tilde{Z}}_{iso-free}(k,\omega) J_0(kr) k dk.$$

The surface slope is thus

$$\frac{\partial}{\partial r}\tilde{Z}_{iso-free}(r,\omega) = \quad (16)$$
$$\frac{1}{2\pi}\int_0^{+\infty} \tilde{\tilde{Z}}_{iso-free}(k,\omega)(-J_1(kr)) k^2 dk.$$

If the surface slope is nearly constant in the range of $r$ covered by the probe beam, the probe beam deflection angle would simply be twice the surface slope. This condition is not well satisfied for a typical experiment because the radii of the probe beam is comparable to the radii of the pump beam. Nevertheless, we previously showed that the following convolution of the probe intensity with the surface slope gives a good description of the probe beam deflection angle [31]:

$$\tilde{\theta}_{iso-free}(r_0,\omega) = \quad (17)$$
$$\frac{C_{probe}}{\pi}\int_0^{+\infty} \tilde{\tilde{Z}}_{iso-free}(k,\omega) e^{-w_0^2 k^2/8}(-J_1(kr_0)) k^2 dk,$$

where $C_{probe}$ is a material-independent constant on the order of unity, $w_0$ the $1/e^2$ radii of pump and probe beam, $r_0$ the offset distance between pump and probe beam.

**B. The full model**

The deformation of the surface of the sample (metal coating/ bulk material) can be different from the prediction of the isotropic free thermal expansion model due to two issues. The first issue is the effects of the metal coating on the deformation the sample surface, including the thermal expansion of the metal coating, the elastic deformation of the bulk material created by the thermal expansion stress of the metal coating, and constraint of the thermal expansion strain of the bulk material by the stiffness of the metal coating. The second issue is the anisotropy of elastic constants and thermal expansion coefficients of the metal coating and the bulk material. Furthermore, the isotropic free thermal expansion of model neglects probe beam deflection created by mechanisms other than surface deformation.

To address these drawbacks of the isotropic free thermal expansion model, we present here "the full model" of probe beam deflection for metal coated bulk material measured in air. The sample surface deformation is calculated for a finite thickness metal coating on a semi-infinite bulk material, where both the metal coating and the bulk material have anisotropic thermal expansion, thermal conductivity, and elastic constants; i.e., thermal expansion and thermal conductivity are described by 2$^{nd}$ rank tensors and the elastic constants are described by 4$^{th}$ rank tensors. We also include the probe beam deflection contributed by the changes of optical path length in air and changes of the phase of the reflection coefficient due to the temperature dependence of the refractive index of air and metal coating.

The Cartesian coordinate system is set up with $x = y = 0$ at the center of the pump beam, $z = 0$ at the surface of the bulk material and $\hat{z}$ pointing inwards. The governing equations of elasticity as expressed by the displacement $\boldsymbol{u} = u_x\hat{x} + u_y\hat{y} + u_z\hat{z}$ are given by [29]

$$\frac{\partial \sigma_{xx}}{\partial x} + \frac{\partial \sigma_{xy}}{\partial y} + \frac{\partial \sigma_{zx}}{\partial z} = \rho \ddot{u}_x, \tag{18a}$$

$$\frac{\partial \sigma_{xy}}{\partial x} + \frac{\partial \sigma_{yy}}{\partial y} + \frac{\partial \sigma_{zy}}{\partial z} = \rho \ddot{u}_y, \tag{18b}$$

$$\frac{\partial \sigma_{zx}}{\partial x} + \frac{\partial \sigma_{zy}}{\partial y} + \frac{\partial \sigma_{zz}}{\partial z} = \rho \ddot{u}_z, \tag{18c}$$

where the stresses are written in terms of displacements by the inverted Duhamel-Neumann law and strain-displacement relation [29],

$$\begin{bmatrix}\sigma_{xx}\\\sigma_{yy}\\\sigma_{zz}\\\sigma_{zy}\\\sigma_{zx}\\\sigma_{xy}\end{bmatrix} \tag{19}$$

$$= \begin{bmatrix} C'_{11} & C'_{12} & C'_{13} & C'_{14} & C'_{15} & C'_{16} \\ C'_{12} & C'_{22} & C'_{23} & C'_{24} & C'_{25} & C'_{26} \\ C'_{13} & C'_{23} & C'_{33} & C'_{34} & C'_{35} & C'_{36} \\ C'_{14} & C'_{24} & C'_{34} & C'_{44} & C'_{45} & C'_{46} \\ C'_{15} & C'_{25} & C'_{35} & C'_{45} & C'_{55} & C'_{56} \\ C'_{16} & C'_{26} & C'_{36} & C'_{46} & C'_{56} & C'_{66} \end{bmatrix} \begin{bmatrix} \frac{\partial u_x}{\partial x} - \alpha_x T \\ \frac{\partial u_y}{\partial y} - \alpha_y T \\ \frac{\partial u_z}{\partial z} - \alpha_z T \\ \frac{\partial u_y}{\partial z} + \frac{\partial u_z}{\partial y} \\ \frac{\partial u_x}{\partial z} + \frac{\partial u_z}{\partial x} \\ \frac{\partial u_x}{\partial y} + \frac{\partial u_y}{\partial x} \end{bmatrix},$$

where $\alpha_x$, $\alpha_y$, and $\alpha_z$ are the linear coefficient of thermal expansion along $\hat{x}$, $\hat{y}$, and $\hat{z}$, respectively (assuming all off-diagonal components in the thermal expansion tensor being zero), the elastic constants $C'_{mn}$ is a component of the forth-rank tensor of stiffness under the current coordinate system (the single number $m$, $n = 1,2,3,4,5,6$ is the abbreviated subscripts of the pairs 11, 22, 33, 23, 31, 12). Since the elastic constants are typically known under the coordinate system of crystal axes, transformation of the tensor of stiffness needs to be performed considering the orientation of the crystal axes with respect to the current coordinate system [32].

The governing equations Eq. (18) are subjected to stress-free condition at the surface of the metal coating and continuity condition at the interface between metal coating and the bulk material as follows [29],

$$\sigma_{zx}|_{z=-L} = 0, \tag{20a}$$
$$\sigma_{zy}|_{z=-L} = 0,$$
$$\sigma_{zz}|_{z=-L} = 0,$$
$$\sigma_{zx}|_{z=0^-} = \sigma_{zx}|_{z=0^+}, \tag{20b}$$
$$\sigma_{zy}|_{z=0^-} = \sigma_{zy}|_{z=0^+},$$
$$\sigma_{zz}|_{z=0^-} = \sigma_{zz}|_{z=0^+},$$
$$u_x|_{z=0^-} = u_x|_{z=0^+}, \tag{20c}$$
$$u_y|_{z=0^-} = u_y|_{z=0^+},$$
$$u_z|_{z=0^-} = u_z|_{z=0^+},$$

where $L$ is the thickness of the metal coating.

With the approximation of the heating by the laser entering from the sample surface, and assuming all off-diagonal components in the thermal conductivity tensor being zero, the temperature field $T$ is governed by:

$$C\dot{T} = \Lambda_x\frac{\partial^2 T}{\partial x^2} + \Lambda_y\frac{\partial^2 T}{\partial y^2} + \Lambda_z\frac{\partial^2 T}{\partial z^2}, \tag{21}$$

After Fourier transforms ($x$ to $\eta$, $y$ to $\xi$, $t$ to $\omega$), the governing equations Eq. (18) and third, fourth, and fifth rows of Eq. (19) are rearranged to the following form [33],

$$A\frac{d\boldsymbol{S}}{dz} = B\boldsymbol{S} + D\hat{T}, \tag{22}$$

$$\boldsymbol{S} = \begin{bmatrix}\hat{u}_x & \hat{u}_y & \hat{u}_z & \hat{\sigma}_{zx} & \hat{\sigma}_{zy} & \hat{\sigma}_{zz}\end{bmatrix}^T,$$

where the matrixes are given by

$$A = \begin{bmatrix} iC'_{15}\eta + iC'_{56}\xi & iC'_{14}\eta + iC'_{46}\xi & iC'_{13}\eta + iC'_{36}\xi & 1 & 0 & 0 \\ iC'_{56}\eta + iC'_{25}\xi & iC'_{46}\eta + iC'_{24}\xi & iC'_{36}\eta + iC'_{23}\xi & 0 & 1 & 0 \\ 0 & 0 & 0 & 0 & 0 & 1 \\ C'_{55} & C'_{45} & C'_{35} & 0 & 0 & 0 \\ C'_{45} & C'_{44} & C'_{34} & 0 & 0 & 0 \\ C'_{35} & C'_{34} & C'_{33} & 0 & 0 & 0 \end{bmatrix}, \quad (23)$$

$$B = \begin{bmatrix} -\rho\omega^2 + C'_{11}\eta^2 + 2C'_{16}\eta\xi + C'_{66}\xi^2 & C'_{16}\eta^2 + (C'_{12} + C'_{66})\eta\xi + C'_{26}\xi^2 & \eta^2 C'_{15} + (C'_{14} + C'_{56})\eta\xi + C'_{46}\xi^2 & 0 & 0 & 0 \\ C'_{16}\eta^2 + (C'_{12} + C'_{66})\eta\xi + C'_{26}\xi^2 & -\rho\omega^2 + C'_{66}\eta^2 + 2C'_{16}\eta\xi + C'_{22}\xi^2 & C'_{56}\eta^2 + (C'_{46} + C'_{25})\eta\xi + C'_{24}\xi^2 & 0 & 0 & 0 \\ 0 & 0 & -\rho\omega^2 & -i\eta & -i\xi & 0 \\ -iC'_{15}\eta - iC'_{56}\xi & -iC'_{56}\eta - iC'_{25}\xi & -iC'_{55}\eta - iC'_{45}\xi & 1 & 0 & 0 \\ -iC'_{14}\eta - iC'_{46}\xi & -iC'_{46}\eta - iC'_{24}\xi & -iC'_{45}\eta - iC'_{44}\xi & 0 & 1 & 0 \\ -iC'_{13}\eta - iC'_{36}\xi & -iC'_{36}\eta - iC'_{23}\xi & -iC'_{35}\eta - iC'_{34}\xi & 0 & 0 & 1 \end{bmatrix}, \quad (24)$$

$$D = \begin{bmatrix} (iC'_{11}\eta + iC'_{16}\xi)\alpha_x + (iC'_{12}\eta + iC'_{26}\xi)\alpha_y + (iC'_{13}\eta + iC'_{36}\xi)\alpha_z \\ (iC'_{16}\eta + iC'_{12}\xi)\alpha_x + (iC'_{26}\eta + iC'_{22}\xi)\alpha_y + (iC'_{36}\eta + iC'_{23}\xi)\alpha_z \\ 0 \\ C'_{15}\alpha_x + C'_{25}\alpha_y + C'_{35}\alpha_z \\ C'_{14}\alpha_x + C'_{24}\alpha_y + C'_{34}\alpha_z \\ C'_{13}\alpha_x + C'_{23}\alpha_y + C'_{33}\alpha_z \end{bmatrix}, \quad (25)$$

The boundary conditions Eq. (20) after Fourier transforms are given by

$$S(3)|_{z=-L} = 0, \quad S(4)|_{z=-L} = 0, \quad (26)$$
$$S(5)|_{z=-L} = 0, \quad S|_{z=0^-} = S|_{z=0^+}$$

The governing equation of temperature field Eq. (21) after Fourier transforms is given by:

$$\frac{\partial^2 \hat{T}}{\partial z^2} = \zeta^2 \hat{T}, \quad (27)$$

$$\zeta = \left(\frac{\Lambda_x}{\Lambda_z}\eta^2 + \frac{\Lambda_y}{\Lambda_z}\xi^2 + \frac{i\omega C}{\Lambda_z}\right)^{1/2},$$

which yields the following solution

$$\hat{T} = \begin{cases} \hat{T}_s e^{\zeta_1(z+L)}, & z < -L \\ a_- e^{\zeta_2 z} + a_+ e^{-\zeta_2 z}, & -L < z < 0, \\ \hat{T}_{bs} e^{-\zeta_3 z}, & z > 0 \end{cases} \quad (28)$$

$$a_- = \frac{(1 + \Lambda_{z,3}\zeta_3/G)\hat{T}_{bs} - \hat{T}_s e^{-\zeta_2 L}}{e^{\zeta_2 L} - e^{-\zeta_2 L}},$$

$$a_+ = \hat{T}_s - a_-,$$

where the subscript of $\Lambda_x$, $\Lambda_z$, $C$, $\zeta$ being 1, 2 and 3 indicating air, the metal coating and the bulk material, respectively, $G$ the thermal boundary conductance of the interface of the metal coating and the bulk material, $\hat{T}_{bs} \equiv \hat{T}|_{z=0}$ and $\hat{T}_s \equiv \hat{T}|_{z=-L}$ are solved by transfer matrix approach for heat conduction in semi-infinite air/ metal coating/ semi-infinite material system as detailed in literature [2, 30].

With temperature field given by Eq. (28), the general solution of the governing equation Eq. (22) is given by

$$-L < z < 0: S(m) = \sum_{n=1}^{6} Q_2(m,n) e^{\lambda_2(n)(z+L)} J_2(n) \quad (29a)$$

$$+ \sum_{n=1}^{6} Q_2(m,n) U_2(n) \left( \frac{a_- e^{\zeta_2(z+L)}}{\zeta_2 - \lambda_2(n)} - \frac{a_+ e^{-\zeta_2(z+L)}}{\zeta_2 + \lambda_2(n)} \right),$$

$$z > 0: S(m) = \sum_{n=1}^{3} Q_3(m,n) e^{\lambda_3(n)(z+L)} J_3(n) \quad (29b)$$

$$- \sum_{n=1}^{6} Q_3(m,n) U_3(n) \frac{\hat{T}_{bs} e^{-\zeta_3 z}}{\zeta_3 + \lambda_3(n)},$$

where $m = 1,2,3,4,5,6$, the vector $J$ is composed of constants to be determined by boundary condition; each column of the six-by-six matrix $Q$ is an eigenvector of $A^{-1}B$, with the corresponding component of the vector $\lambda$ being the eigenvalue to which the eigenvector belongs; the vector $U = Q^{-1}AD$. The subscript of $Q$, $\lambda$, $U$ and $J$ being 2 and 3 indicating the metal coating and the bulk material, respectively. Note that $Q_3$ and $\lambda_3$ are arranged such that the real part of the first three components of $\lambda_3$ are negative, representing wave propagation along the positive $z$ direction. To make the determination of the sign of components of $\lambda_3$ free from influence of numerical error, we adopt the strategy of including a small imaginary part in the elastic constants, e.g., by multiplying $(1 + i \times 10^{-4})$.

To find the sample surface displacement $\hat{Z}_{ani}(\eta,\xi,\omega) \equiv -\hat{u}_z|_{z=-L} = -S(3)|_{z=-L}$, the constants $J_2$ are determined by solving the linear equations resulting from substitution of the general solution Eq. (29) for $S$ in the boundary conditions Eq. (26). The sample surface displacement is given by,

$$\hat{Z}_{ani}(\eta,\xi,\omega) = -\sum_{n=1}^{6} Q_2(3,n) J_2(n) \quad (30)$$

$$- \sum_{n=1}^{6} Q_2(3,n) U_2(n) \left( \frac{a_-}{\zeta_2 - \lambda_2(n)} - \frac{a_+}{\zeta_2 + \lambda_2(n)} \right).$$

The probe beam deflection due to changes of optical path length with temperature can regarded as resulting from an effective surface displacement [26]. The effective surface

displacement due to changes of optical path length in air with temperature is given by [26]

$$\hat{Z}_{air}(\eta, \xi, \omega) = -\frac{dn_{air}}{dT} \frac{\hat{T}_s}{\zeta_1}, \quad (31)$$

where $\frac{dn_{air}}{dT}$ is the temperature derivative of refractive index of air. Besides, at the air/metal coating interface, changes of the phase of the Fresnel reflection coefficient $\phi_r$ with temperature gives the effective surface displacement [26],

$$\hat{Z}_r(\eta, \xi, \omega) = -\frac{\lambda_1}{4\pi} \frac{d\phi_r}{dT} \hat{T}_s, \quad (32)$$

where $\lambda_1$ is wavelength of the probe beam, $\frac{d\phi_r}{dT}$ the temperature derivative of the phase of the Fresnel reflection coefficient, a constant determined by

$$\phi_r = \pi + \tan^{-1}\left(\frac{\kappa_{metal}}{n_{air} - n_{metal}}\right) \quad (33)$$
$$- \tan^{-1}\left(\frac{\kappa_{metal}}{n_{air} + n_{metal}}\right),$$

where $n_{metal}$ and $\kappa_{metal}$ are the real part and imaginary part of the refractive index of the metal coating, respectively. Eq. (33) assumes that the metal coating is optically semi-infinite for the probe beam.

The probe beam deflection contributed due to each surface displacement $\hat{Z}_i(\eta, \xi, \omega)$ (where $i$ refers to $ani$, $air$, or $r$) is then calculated in a similar way as described for the isotropic free expansion model. Inverse Fourier transforms are applied to obtain the surface displacement in real space,

$$\tilde{Z}_i(x, y, \omega) = \quad (34)$$
$$\frac{1}{(2\pi)^2} \int_0^{+\infty} \int_0^{+\infty} \hat{Z}_i(\eta, \xi, \omega) e^{i(\eta x + \xi y)} d\xi \, d\eta,$$

By making the transforms of variables

$$x = r\cos(\varphi), \quad y = r\sin(\varphi), \quad (35a)$$
$$\eta = k\cos(\psi), \quad \xi = k\sin(\psi), \quad (35b)$$

the surface slope is found to be

$$\frac{\partial}{\partial r} \tilde{Z}_i(r, \varphi, \omega) = \frac{1}{(2\pi)^2} \times \quad (36)$$
$$\int_0^{+\infty} k^2 dk \int_0^{2\pi} \hat{Z}_i(k, \psi, \omega) i\cos(\psi - \varphi) e^{ikr\cos(\psi - \varphi)} d\psi$$

Following the approximation of convolution of the probe intensity with the surface slope, the probe beam deflection angle is given by

$$\tilde{\theta}_i(r_0, \varphi_0, \omega) = \frac{C_{probe}}{2\pi^2} \times \quad (37)$$
$$\int_0^{+\infty} e^{-w_0^2 k^2/8} k^2 dk \int_0^{2\pi} \hat{Z}_i(k, \psi, \omega) i\cos(\psi - \varphi) e^{ikr\cos(\psi - \varphi)} d\psi,$$

where $C_{probe}$ is a material-independent on the order of unity, $w_0$ the $1/e^2$ radii of probe beam, $r_0$ the offset distance between pump and probe beam, $\varphi_0$ the angle of the vector from the center of pump beam to that of the probe beam in the current coordinate system. The MATLAB codes for implementation of the above models are available for download at:
https://zenodo.org/badge/latestdoi/529379931.

## III. MEASUREMENT OF PROBE BEAM DEFLECTION

The optics of the frequency-domain probe-beam-deflection (FD-PBD) system are shown in Figure 1b. The only significant difference between our FD-PBD system and a typical FDTR measurement system is our use of a quadrant-cell photoreceiver for detecting modulations in beam displacement in place of the single photodiode or balanced photoreceiver for detecting modulation of beam intensity. Two CW diode lasers at wavelength of 660 nm and 640 nm are used as the pump beam and probe beam, respectively. The two laser beams are focused by the objective lens and incident on the Al coated bulk material at normal incidence with a beam offset along the vertical direction. With the pump beam modulated at frequency $f$, the heating due to absorption by the Al coating creates elastic deformation of the sample surface and consequently an angular deflection of the probe beam along the vertical direction synchronous with the modulation frequency. After the reflected probe beam passes back through the objective lens, this angular deflection is converted to a transverse displacement along the vertical direction, which is measured by the difference signal (between the upper cells and lower cells) of the quadrant-cell photoreceiver which has been aligned with the probe beam. With this signal as the input and the modulation signal of pump beam as the reference, a lock-in amplifier yields the in-phase signal $X$ and out-of-phase signal $Y$ at frequency $f$.

In fact, due to the finite bandwidth of the quadrant-cell photoreceiver, the directly measured signal contains nontrivial contribution from quadrant-cell photoreceiver at high frequencies (higher than 1 kHz for the Newport[TM] 2901 quadrant-cell photoreceiver). To extract the PBD signal, we divide the complex signal $X + iY$ at each $f$ by the response of the quadrant-cell photoreceiver at the same $f$. We obtain the response of the photoreceiver by collecting the complex signal with the pump beam off and probe beam modulated, and then normalizing it by its amplitude in the low frequency limit.

To extend the measurements to frequencies $f$ above the frequency response of the quadrant-cell photoreceiver, we modulate the pump at frequency $f$, modulate the probe beam at frequency $f + f_{dec}$, and detect the signal at the difference frequency $f_{dec}$. We typically set $f_{dec} = 5$ kHz. The reference signal for the lock-in amplifier is generated by inputting the two modulation signals into a frequency mixer (Mini-Circuit[®] ZAD-6+) and extracting the difference signal at frequency $f_{dec}$ with a lowpass filter (Thorlabs EF114).

The basic procedure of the experiment is as follows. First, the pump and probe beam are focused on the surface of the sample by moving the sample on the translation stage and monitoring the formation of a sharp image of defects or dusts on the sample surface using the video camara. Second, the quadrant-cell photoreceiver is aligned with the probe beam by adjusting the transverse position of the photoreceiver until the dc sum signal of all cells is a maximum and the dc difference signal between the upper cells and lower cells is zero. Third, the pump beam is aligned with the probe beam by rotating the beam-splitter near the objective lens until the two beam spots overlap as seen from the video camera and the lock-in amplifier yields zero signal. Fourth, a specified beam offset is set by rotating the beam-splitter according to the micrometer. The proportional factor between the beam offset distance and reading of the micrometer has been measured before the experiment. For 10x microscope objective (focal length of 20 mm) and the Newport[TM] SL8A gimbal mount, the calibration in vertical direction is 4.9 μm beam offset per 10 mm reading on the differential micrometer of the precision gimbal. Finally, the in-phase and out-of-phase signal at a series of modulation frequencies are collected from the lock-in amplifier.

## IV. RESULTS AND DISCUSSION

### A. Analysis of calculated PBD signal

As shown in Fig. 2, the probe beam deflection angles (PBD) are calculated using the full model (Eq. 37 with $C_{probe} = 1$) for four representative samples (polystyrene, SrTiO$_3$, MoS$_2$, and Au) measured in air. The real part and imaginary part of the calculated complex number $\tilde{\theta}_i$ is the in-phase and out-of-phase PBD (in unit of μrad), respectively. Three sources of PBD are present, i.e., surface deformation (in red), changes of optical path length in air (in blue), and changes of the phase of the reflection coefficient (in green). Fig. 2a shows PBD for polystyrene, an amorphous polymer with a low thermal conductivity of 0.156 W/(m K). Due to the high coefficient of thermal expansion $\alpha_T = 73\times10^{-6}$ K$^{-1}$, the PBD due to surface deformation completely dominates the total PBD. Fig. 2b shows PBD for cubic SrTiO$_3$ (100), an oxide crystal with a thermal conductivity of 11.0 W/(m K). Due to the in-plane anisotropy of elastic constants, the PBD due to surface deformation is dependent on the in-plane orientation of beam offset. It is assumed here that the beam offset is along [010]. With $\alpha_T = 10.4\times10^{-6}$ K$^{-1}$, the PBD due to surface deformation still dominants the total PBD. Fig. 2c is for bulk hexagonal MoS$_2$, a material with highly anisotropic thermal expansion, thermal conductivity, and elastic constants. Due to the low coefficient of thermal expansion along the radial direction ($5.8\times10^{-6}$ K$^{-1}$), PBD due to surface deformation is no longer large enough to dominate the total PBD, with the PBD due to air also playing an important role. Fig. 2d shows PBD for Au, a cubic crystal metal with a high thermal conductivity of 315 W/(m K). As is the case for most metals, Au has an optical absorption depth that is much smaller than the laser spot size. Au is highly reflective but has sufficient optical absorption at the wavelength of pump beam (660 nm) to absorb a significant fraction of the probe light and thus removes the need for the Al coating. Calculation of PBD for bare metal can be done using the same formalism as described above with the metal coating set to the same material as the bulk sample.

To gain insight on the dependence of the PBD signal on materials parameters, the frequency and value of the peak of the out-of-phase PBD are listed in Table I, together with the thermal diffusivity and coefficient of thermal expansion. The peak frequency of the PBD due to surface deformation scales, to a good approximation, with the in-plane thermal

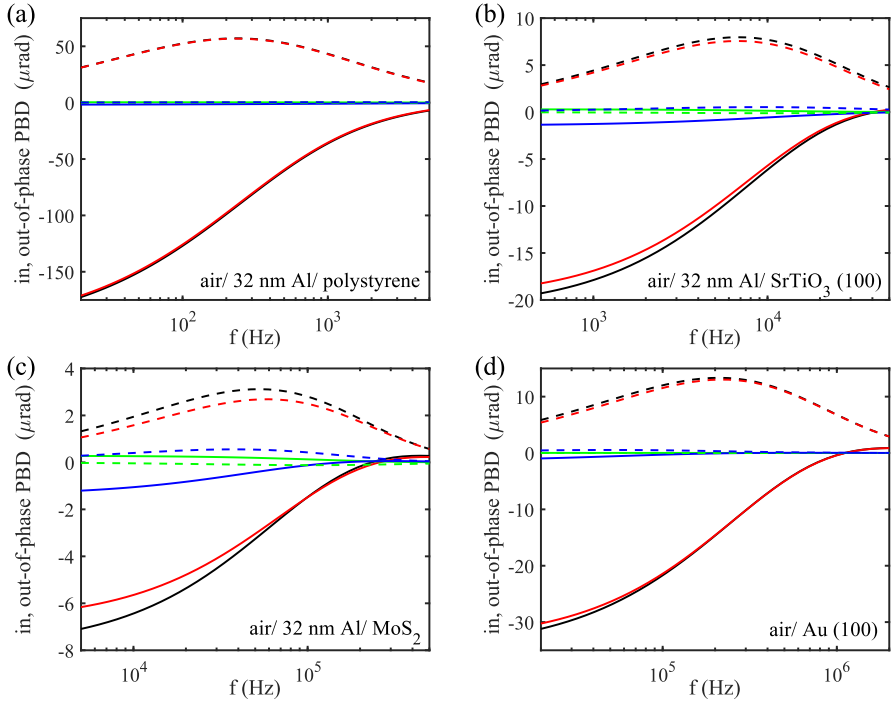

**FIG. 2.** Calculated in-phase (solid lines) and out-of-phase (dashed lines) PBD in air: (a) 32 nm Al coated polystyrene, (b) 32 nm Al coated SrTiO$_3$ (100) with beam offset along [010], (c) 32 nm Al coated MoS$_2$ (001), and (d) bare Au (100) with beam offset along [010]. Red, blue, green, and black curves represent PBD due to surface deformation, air, reflection coefficient, and the total, respectively. Pump and probe beam $1/e^2$ radii $w_0 = 8.3$ μm, beam offset $r_0 = 19.6$ μm. The heating power is chosen for each sample such that the maximum amplitude of surface temperature rise equals 3 K.

TABLE I. The frequency and value of the peak of out-of-phase PBD in Fig.2 and materials parameters ($w_0 = 8.3$ μm, $r_0 = 19.6$

|  | polystyrene |  | SrTiO$_3$ |  | MoS$_2$ |  | Au |  |
| --- | --- | --- | --- | --- | --- | --- | --- | --- |
|  | $f$ (kHz) | PBD (μrad) | $f$ (kHz) | PBD (μrad) | $f$ (kHz) | PBD (μrad) | $f$ (kHz) | PBD (μrad) |
| Surface deformation | 0.23 | 56.79 | 6.7 | 7.57 | 60 | 2.69 | 212 | 13.02 |
| Reflection coefficient | 2.36 | -0.09 | 10.8 | -0.12 | 108 | -0.12 | 302 | -0.01 |
| Air | 1.63 | 0.41 | 9.6 | 0.53 | 37 | 0.56 | 51 | 0.54 |
| Total | 0.23 | 57.06 | 6.7 | 7.97 | 53 | 3.12 | 212 | 13.34 |
| $D = \Lambda/C$ (×10$^{-6}$ m$^2$/s) | 0.12[α] | | 4.0[γ] | | in/through-plane 42/2.5[ε] | | 127[η] | |
| $\alpha_T$ (e-6/K) | 73[β] | | 10.4[δ] | | in/through-plane 5.8/11.3[ζ] | | 14.2[η] | |

[α] L.C.K. Carwile and H.J. Hoge, (1966).
[β] Z. Zhang, P. Zhao, P. Lin, and F. Sun, Polymer (Guildf) **47**, 4893 (2006).
[γ] Y. Suemune, J Physical Soc Japan **20**, 174 (1965).
[δ] D. de Ligny and P. Richet, Phys Rev B **53**, 3013 (1996).
[ε] E. López-Honorato, C. Chiritescu, P. Xiao, D.G. Cahill, G. Marsh, and T.J. Abram, Journal of Nuclear Materials **378**, 35 (2008).
[ζ] C.K. Gan and Y.Y.F. Liu, Phys Rev B **94**, 134303 (2016).
[η] W.M. Haynes, D.R. Lide, and T.J. Bruno, *CRC Handbook of Chemistry and Physics* (CRC press, 2016).

diffusivity of the bulk material. This makes it possible to measure the thermal conductivity without knowing the proportional factors that only influence that magnitude of the signal, for example the incident laser power and coefficient of optical absorption. In addition, as expected, the peak value of PBD due to air and reflection coefficient do not vary significantly with the type of material under investigation, while PBD due to surface deformation scales with the coefficient of thermal expansion. This observation motivates the following estimate of the magnitude of the PBD angle due to surface deformation,

$$\theta \approx \alpha_T \Delta T, \quad (38)$$

where $\Delta T$ is the amplitude of surface temperature rise.

To check Eq. (38), we note that $\Delta T = 3$ K in the low frequency limit (for each sample, the heating power has been chosen such that the amplitude of surface temperature oscillation in the low frequency limit is equal to 3 K), and therefore Eq. (38) gives an estimate of 220, 31, 17, and 43 μrad for polystyrene, SrTiO$_3$, MoS$_2$, and Au, respectively. These estimates compare well with the calculations of the full model of 174, 18, 6, and 31 μrad shown in Fig. 2.

To compare the magnitude of the signal measured by the FD-PBD system with that of FDTR, the probe beam deflection $\theta$ is converted to a ratio of PBD signal over the dc signal of the photoreceiver as follows. After the objective lens, the transverse displacement of the unfocused probe beam is

$$d = \theta f_0, \quad (39)$$

where $f_0$ is the focal length of the objective. The intensity profile of the unfocused gaussian probe beam is given by

$$I(r) = I_1 e^{-2r^2/w^2}, I_1 = \frac{2P_1}{\pi w^2}, w = \frac{\lambda_1 f_0}{\pi w_0} \quad (40)$$

where $P_1$ is the total power, $I_1$ the peak intensity, $w$ and $w_0$ the 1/e$^2$ radii of unfocused and focused probe beam, respectively. With $d \ll w$, the ratio of interest, i.e., the difference of power (between the upper cells and lower cells) over the total power is given by

$$\left(\frac{8}{\pi}\right)^{1/2} \frac{d}{w} = (8\pi)^{1/2} \frac{w_0}{\lambda_1} \theta = 65.0\theta, \quad (41)$$

given $\lambda_1 = 640$ nm, $w_0 = 8.3$ μm. Using Eq. (38), this ratio, i.e., the normalized PBD signal, is estimated to be $65.0\alpha_T \Delta T$. For most materials, $\alpha_T$ is larger than $5\times 10^{-6}$ K$^{-1}$, which gives a normalized PBD signal $> 3\times 10^{-4}$ K$^{-1}\Delta T$. In comparison, the corresponding value in FDTR measurement is given by $\frac{1}{R}\frac{dR}{dT}\Delta T$. Among the common combination of metal coating and wavelength of laser, Au at wavelength of 532 nm gives the maximum value of $\frac{1}{R}\frac{dR}{dT}\Delta T = 3\times 10^{-4}$ K$^{-1}\Delta T$ [20]. Thus, the magnitude of PBD signal is generally larger than that of FDTR, and more importantly, independent of the choice of metal coating and laser wavelength.

### B. Applicability of the isotropic free thermal expansion model

For extracting thermal conductivity by fitting to the measured PBD data, the isotropic free thermal expansion model is preferred compared to the full model, as the elastic constants and thermal expansion coefficient of the bulk material are not needed. This can be seen from Eq. (14) and Eq. (17), where the Poisson ratio and coefficient of thermal expansion coefficient only enter the results as proportional factors. To have a sense of the applicability of the isotropic free thermal expansion model, its error is evaluated by fitting it to the PBD data generated by the full model and then monitoring the error of the fitted thermal conductivity, as shown in Fig. 3. Due to the symmetry of crystal structure, the PBD and consequently the fitted thermal conductivity of c-plane sapphire, MoS$_2$ (001), and Bi$_2$Se$_3$ (001) are independent of the in-plane orientation of beam offset. This is not the case for the cubic crystal, for which a bar is used to indicate the range of fitted thermal conductivity with a circle indicating the average of these fitted values.

In this analysis, the materials are assumed to be measured in vacuum to avoid the contribution coming from the PBD signal of temperature field in the air above the sample. Otherwise, the fact that the simplified model neglects the PBD contribution from air would have caused noticeable error for materials with relatively small coefficient of thermal expansion, such as borosilicate glass and $MoS_2$. Furthermore, we set the Al coating thickness to 32 nm to reduce the effects of the metal coating on the deformation the sample surface while ensuring that less than 1% of the pump laser power enters the bulk material.

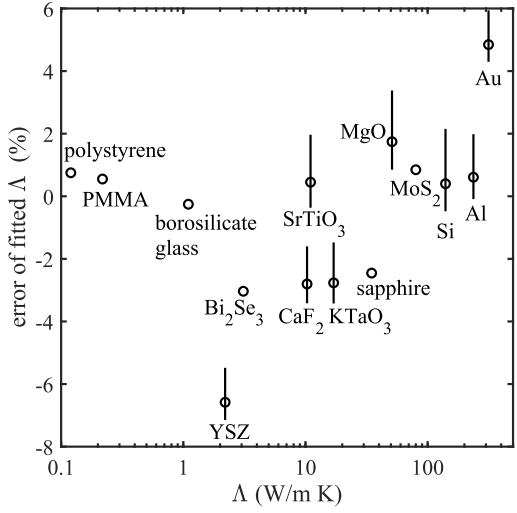

**FIG. 3.** Calculated error of thermal conductivity fitted using the isotropic free thermal expansion model. The full model is used to generate the PBD data for fitting. The sapphire has c-plane (0001) surface orientation, hexagonal crystal ($MoS_2$) and trigonal crystal ($Bi_2Se_3$) have (001) orientation. Cubic crystals (YSZ, $SrTiO_3$, $CaF_2$, $KTaO_3$, MgO, Si, Al, and Au) have (100) orientation, for which the bars indicate the range of fitted thermal conductivity for all possible in-plane orientations of beam offset while the circles indicate the average of the fitted values. $w_0 = 8.3$ μm, $r_0 = 19.6$ μm.

The error inherent in the simplified model is less than ~6% for this selection of materials and measurement conditions. Therefore, by using a relatively thin metal coating and carrying out the measurement in vacuum, the isotropic free thermal expansion model is widely applicable. A further implication of this error analysis is that when using the full model, uncertainties in the elastic constants and thermal expansion coefficient have only a small effect on the accuracy of the fitted thermal conductivity.

### C. Sensitivity and uncertainty

To quantify the sensitivity of the PBD angle to the parameters in the model, we calculate the sensitivity of in-phase PBD angle, $\text{Re}(\tilde{\theta}_{iso-free})$, and out-of-phase PBD angle, $\text{Im}(\tilde{\theta}_{iso-free})$, to a parameter $x$ in the isotropic free thermal expansion model (Eq. 17) as defined by

$$S_{in} = \frac{d\,\text{Re}(\tilde{\theta}_{iso-free})/In_{max}}{dx/x}, \quad (42a)$$

$$S_{out} = \frac{d\,\text{Im}(\tilde{\theta}_{iso-free})/Out_{max}}{dx/x}, \quad (42b)$$

where $In_{max}$ and $Out_{max}$ are the maximum magnitude of in-phase PBD angle and out-of-phase PBD angle in the frequency range studied. For an Al coated $SrTiO_3$ sample, as shown in Fig. 4, $S_{in}$ and $S_{out}$ are calculated for parameter $x$ as follows: in-plane thermal conductivity $\Lambda_r$, through-plane thermal conductivity $\Lambda_z$, volumetric heat capacity $C$ of SrTiO$_3$; thickness $L_{Al}$, thermal conductivity $\Lambda_{Al}$, and volumetric heat capacity $C_{Al}$ of Al; thermal boundary conductance of Al/SrTiO$_3$ interface $G$; pump and probe beam radii $w_0$; and the beam offset $r_0$. By comparison of the first row (Fig 4a and 4b) with $w_0 = 8.3$ μm, $r_0 = 19.6$ μm and the second row (Fig 4c and 4d) with $w_0 = 8.3$ μm, $r_0 = 9.8$ μm, it is seen that with higher ratio of $r_0/w_0$, sensitivity to $w_0$ is lower while sensitivity to $r_0$ is higher. Since the beam offset $r_0$ can be measured at a higher accuracy than the beam radii $w_0$, $w_0 = 8.3$ μm, $r_0 = 19.6$ μm should reduce the total uncertainty due to uncertainty of parameters.

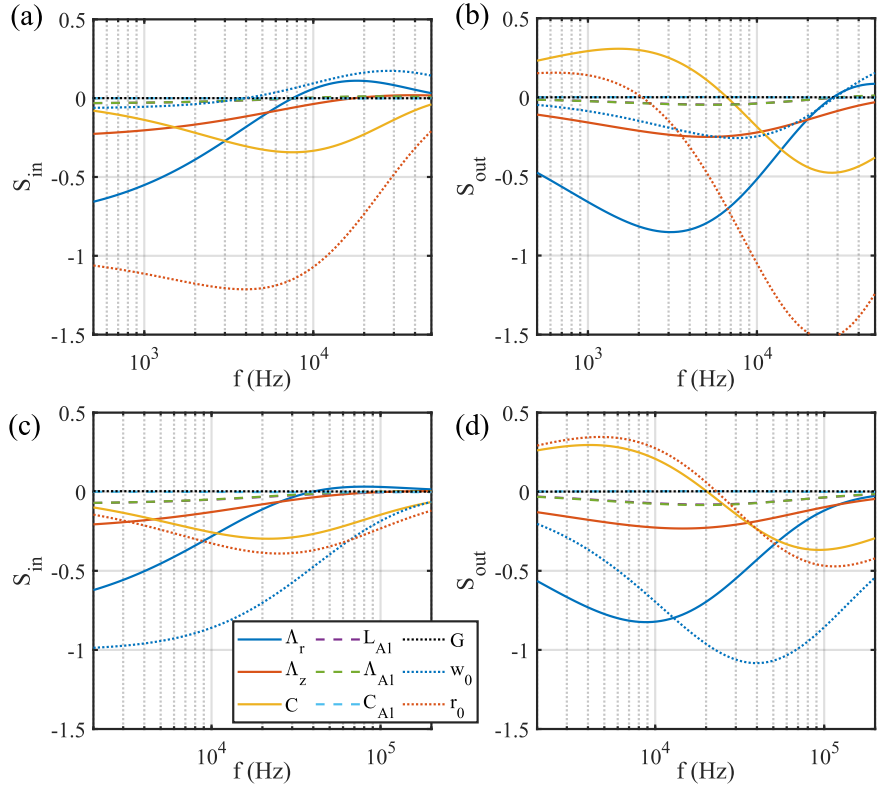

**FIG. 4.** Calculated sensitivity and PBD of 32 nm Al coated $SrTiO_3$ (100) in vacuum using the isotropic free thermal expansion model: (a) (b) with $w_0 = 8.3$ μm, $r_0 = 19.6$ μm; (c) (d) with $w_0 = 8.3$ μm, $r_0 = 9.8$ μm.

Note that for SrTiO$_3$, $\Lambda_r = \Lambda_z$, but the sensitivity to $\Lambda_r$ is significantly larger than the sensitivity to $\Lambda_z$ at most frequencies. The reason is that PBD is primarily dependent on the radial derivative of temperature and is thus primarily sensitivity to the in-plane thermal diffusion. Partly due to the same reason, sensitivity to the thermal boundary conductance $G$ is the nearly zero. The presumably dominate reason of low sensitivity to $G$ is the large thermal penetration depth ~ $\sqrt{D/f}$ (20 μm at $f = 10$ kHz). Besides, it is noticed that the sensitivity to the heat capacity $C$ significantly deviate from -1 times the sensitivity to the thermal conductivity $\Lambda_r$, the expectation behavior if $\Lambda_r$ and $C$ enter the results merely in the form of in-plane thermal diffusivity $\Lambda_r/C$.

To quantify the uncertainty of measured thermal conductivity due to the parameters, uncertainty due to each parameter $x$ is calculated as follows. The PBD data is first calculated using the isotropic free thermal expansion model with the value of $x$ increased by its uncertainty. The thermal conductivity is then extracted from this PBD data by fitting using the same model with the normal value of $x$. The uncertainty of thermal conductivity due to uncertainty of $x$ is evaluated as the error of the fitted thermal conductivity. With the values and uncertainties of parameters in Table II, the calculated uncertainties of thermal conductivity are shown in Fig. 5. Three different ways of fitting are considered, i.e., fitting to both in-phase and out-of-phase PBD (lines), only in-phase PBD (open circles), and only out-of-phase PBD (filled circles). The total uncertainty (in black) is the square root of the sum of square of uncertainties of all parameters. Fitting to both in-phase and out-of-phase PBD gives a total uncertainty roughly constant at 5% when thermal conductivity varies within 0.1 W/(m K) - 300 W/(m K). The uncertainty of the beam offset $r_0$ is identified as the dominant source of the total uncertainty, due to the high sensitivity to $r_0$ (with $w_0 = 8.3$ μm, $r_0 = 19.6$ μm used). If $w_0 = 8.3$ μm, $r_0 = 9.8$ μm are used instead, uncertainty of $w_0$ will become the dominant source of uncertainty and result in a larger total uncertainty due to the higher uncertainty of $w_0$ than $r_0$ (see Table II).

### D. Fitting to measured PBD signal

As providing a real-world example, we plot as Fig. 6a the best fit of the measured PBD signal of Al coated SrTiO$_3$ to the isotropic free thermal expansion model with the thermal conductivity as the only fitting parameters (during the fitting, in-phase and out-of-phase PBD signal are first divided by their respective maximum magnitude in the frequency range). In total 50 data points are collected over frequency range of a factor of 100 from a frequency 10 times small than the peak in the out-of-phase signal to a frequency 10 times larger than the peak. Nearly perfect agreement between the fitting curve and the measured signal is found, yielding the thermal conductivity of 10.4 W/(m K), only 5% smaller than the accepted value of 11.0 W/(m K).

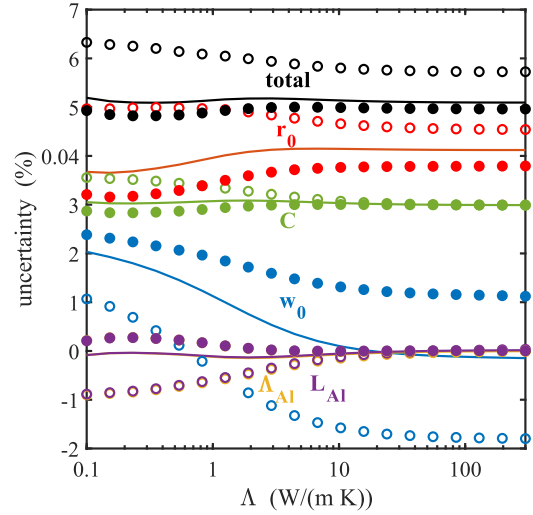

**FIG. 5.** Calculated uncertainty of thermal conductivity due to uncertainty of input parameters using the isotropic free thermal expansion model. Lines, open circles, and filled circles represent fitting to both in-phase and out-of-phase, only in-phase, and only out-of-phase PBD. Each color indicates uncertainty due to one source while the black indicating the total uncertainty. Uncertainties due to $C_{Al}$ and $G$ are negligible and omitted from the plot. Values of input parameters and their uncertainties are listed in Table II.

TABLE II. Parameters and uncertainties for uncertainty analysis.

| $C$ | $\Lambda_{Al}$ | $C_{Al}$ | $L_{Al}$ | $w_0$ | $r_0$ | $G$ |
|---|---|---|---|---|---|---|
| 2 | 165 | 2.42 | 32 | 8.3 | 19.6 | 40 |
| MJ/(m$^3$K) | W/(m K) | MJ/(m$^3$K) | nm | μm | μm | MW/(m$^2$K) |
| 3% | 5% | 3% | 5% | 8% | 2% | 50% |

The normalized PBD signal given by Eq. (41) is equal to the ratio of amplitude of the PBD signal, $V$, over the dc sum signal of all cells of the quadrant-cell photoreceiver, SUM, thus giving the relation,

$$V/SUM = 65.0\theta, \qquad (43)$$

The experimental data in Fig. 6a is measured with SUM = 0.842 V. Therefore, a typical magnitude of PBD signal ~ 100 μV in Fig. 6a corresponds a PBD angle $\theta$ ~ 2 μrad. This signal is accompanied by a steady state temperature rise due to pump laser ~ 1.0 K and $\Delta T$ ~ 0.8 K at the lower end of the frequency range (as calculated using the measured absorbed power of pump laser $P_0 = 0.36$ mW). With the time constant of the lock-in amplifier set to 1s, data collection takes 5 s at each frequency and 250 s in total, while at $f > 4$ kHz, noise of in-phase and out-of-phase signal (as measured by the standard deviation) is ~ 0.1 μV, i.e., ~ 2 nrad according to Eq. (43). This noise is equal to that measured at the limit of the power of probe beam approaching zero, and is supposed to originate from the quadrant-cell photoreceiver and the lock-in amplifier. At $f$ < 1 kHz, the noise scales as $1/f$ and increases with the power of probe beam. It is thus seen that PBD signal has excellent signal-to-noise ratio and is suitable for high throughput measurement.

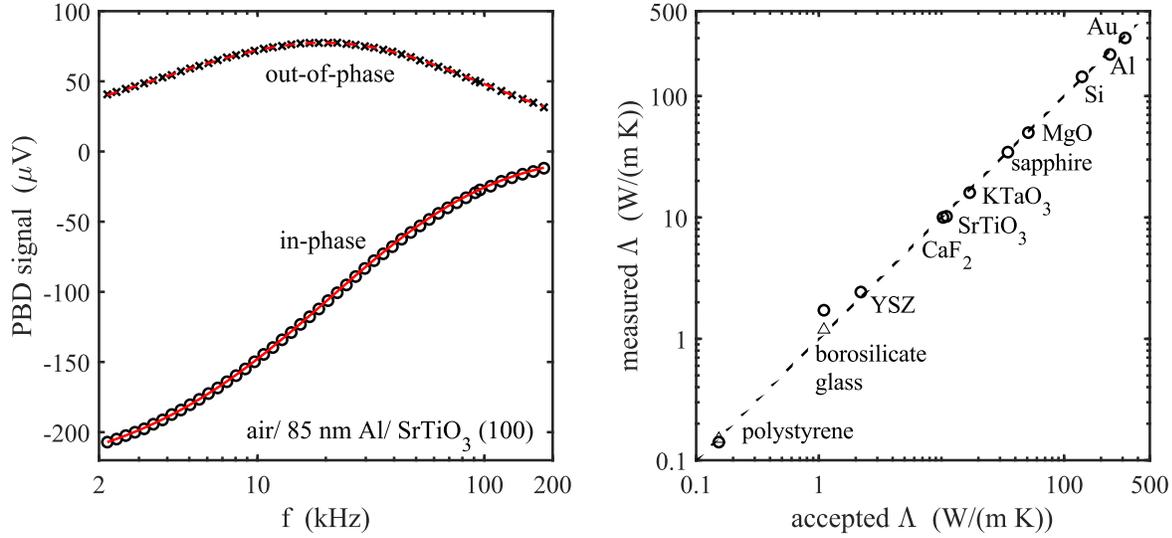

**FIG. 6.** (a) PBD signal of 85 nm Al coated SrTiO$_3$ measured in air and the best fitting using the isotropic free thermal expansion model with the thermal conductivity as the free parameter. (b) Comparison of the measured thermal conductivity using the isotropic free thermal expansion model (circles) and the full model (triangles, only for polystyrene and borosilicate glass) with accepted values for selected materials. All materials are measured in air with ~85 nm Al coating (except Al and Au, which have no metal coating), $w_0$ = 8.3 μm, $r_0$ = 9.8 μm.

In Fig. 6b, we compare thermal conductivities measured by FD-PBD with accepted values for selected materials that span a factor of 3000 in thermal conductivity Measured thermal conductivities agree well with the accepted values except for borosilicate glass. Therefore, the full model is used to fit the same set of measured PBD data for borosilicate glass and the accuracy of the measurement is greatly improved as shown by the triangle for borosilicate. We attribute the relatively large error of the isotropic free thermal expansion model for the borosilicate glass sample to the combination of the small thermal expansion coefficient, the contribution of air to the PBD signal, and the significant effects of the relatively thick Al film (87 nm) on the surface deformation. We also checked the full model for fitting the data for polystyrene. The full model gives essentially the same value of thermal conductivity as the isotropic free thermal expansion model. As an indicator of the overall accuracy of the measurement, the root mean square of the relative deviation of the measured values from the accepted values is 5.7%.

## V. CONCLUSION

In summary, we developed a frequency-domain probe beam deflection (FD-PBD) method for measurement of thermal conductivity of bulk materials. We derived a simplified model for probe beam deflection, the isotropic free thermal expansion model, which allows measurement of thermal conductivity without knowing elastic and thermal expansion properties of material. This simplified model is shown to be widely applicable with small error for measurement with relatively thin metal coating in vacuum. We also derived the full model the incorporates the anisotropic elastic constants and thermal expansion coefficient of the material that is under investigation. We established two unique features of this method as compared to FDTR: 1) a larger signal independent of type of metal coating and laser wavelength; 2) reduced sensitivity to pump and probe beam radii with increasing beam offset. Finally, we validated this method by the matching the measured signal and model prediction, and evaluating the agreement of measured thermal conductivities with accepted values in the range of 0.1 - 300 W/(m K).

## ACKNOWLEDGMENTS

This research is supported by the NSF through the University of Illinois at Urbana-Champaign Materials Research Science and Engineering Center DMR-1720633. The authors declare no conflict of interest.

## AUTHOR DECLARATIONS
### Conflict of Interest

All authors declare that they have no conflicts of interest.

### DATA AVAILABILITY

The data supporting findings of this study are available from the corresponding author upon reasonable request.